# Dynamic Cross-Layer Beamforming in Hybrid Powered Communication Systems With Harvest-Use-Trade Strategy

Yanjie Dong, *Student Member, IEEE*, Md. Jahangir Hossain, *Member, IEEE*, Julian Cheng, *Senior Member, IEEE*, and Victor C.M. Leung, *Fellow, IEEE*

*Abstract*—The application of renewable energy is a promising solution to realize the *Green Communications*. However, if the cellular systems are solely powered by the renewable energy, the weather dependence of the renewable energy arrival makes the systems unstable. On the other hand, the proliferation of the smart grid facilitates the loads with two-way energy trading capability. Hence, a hybrid powered cellular system, which combines the smart grid with the base stations, can reduce the grid energy expenditure and improve the utilization efficiency of the renewable energy. In this paper, the long-term grid energy expenditure minimization problem is formulated as a stochastic optimization model. By leveraging the stochastic optimization theory, we reformulate the stochastic optimization problem as a per-frame grid energy plus weighted penalized packet rate minimization problem, which is NP-hard. As a result, two suboptimal algorithms, which jointly consider the effects of the channel quality and the packet reception failure, are proposed based on the successive approximation beamforming (SABF) technique and the zero-forcing beamforming (ZFBF) technique. The convergence properties of the proposed suboptimal algorithms are established, and the corresponding computational complexities are analyzed. Simulation results show that the proposed SABF algorithm outperforms the ZFBF algorithm in both grid energy expenditure and packet delay. By tuning a control parameter, the grid energy expenditure can be traded for the packet delay under the proposed stochastic optimization model.

## I. Introduction

The wireless data traffic will exceed 700 exabytes in 2016, and this amount will increase nearly threefold over the next 5 years [1]. The wireless data traffic requires a drastic increase in energy consumption, which translates to a significant amount of greenhouse gas emissions and a surge of the energy bill [2]. These emerging issues motivate the information and communication technology sector to search for the green communication solutions. The base stations (BSs) account for 70% to 80% of the overall energy bill of a typical wireless communication system (WCS) [2]–[4], and tremendous number of BSs will be deployed by 2020 to cope with the target of 1000X objective [5]. Therefore, a promising way to reduce the energy bill of the WCSs is to facilitate the application of the renewable energy resources, such as solar radiation and wind energy, at the BSs [6].

As an ecology and economy friendly solution, energy harvesting (EH) technology has been intensively researched for the past several decades [6]–[9]. Equipped with energy harvesters, the BSs can harvest energy from the renewable energy resources. As a result, the high grid energy expenditure caused by the huge traffic demand is reduced [6], [10]–[12]. Nevertheless, the renewable energy resources are highly weather dependent and space-varying [6]; therefore, it is impossible to maintain an acceptable communication quality-of-service (QoS) when the BSs are solely powered by the renewable energy resources. Hence, a hybrid powered communication (HPC) system is preferred and it is therefore the focus of this paper.

In the HPC systems, multiple energy usage strategies are proposed to deal with the uncertainty of the renewable energy resources, such as harvest-store-use (HSU) strategy [10]–[12] and harvest-use-store (HUS) strategy [13], [14]. In the first strategy, the BS stores the harvested energy into a storage medium before usage. Unlike the HSU strategy, the HUS strategy prioritizes the usage of the harvested energy and the surplus is stored in the storage medium. Yet, both the HSU and the HUS strategies require the energy storage medium, which suffers from imperfections such as storage loss and energy leakage [8], [13]–[15]. The authors in [15] investigated the long-term grid energy minimization problem in the HPC systems with HUS strategy. However, the two-way energy trading, which can help the cellular operators generate extra revenue, is ignored. On the other hand, the traditional electricity grid is experiencing a paradigm shift to a smart grid, where the two-way energy trading between the loads and the grid becomes applicable [16]. With the two-way trading capability in the HPC systems, a harvest-use-trade (HUT) strategy is preferred to thoroughly resolve the imperfections of the storage medium, which can improve the utilization efficiency of the harvested energy [17].

### A. Related Works and Motivations

There are two major topics in the HPC systems, namely minimizing the grid energy expenditure under certain performance requirements [14]–[24] and improving the system performance subject to short-term and/or long-term power constraints [11], [18], [23]–[27]. Several research efforts have

Yanjie Dong and Victor C.M. Leung are with the Department of Electrical and Computer Engineering, The University of British Columbia, Vancouver, BC, V6T 1Z4 Canada (email:{ydong16, vleung}@ece.ubc.ca).

Md. Jahangir Hossain and Julian Cheng are with the School of Engineering, The University of British Columbia, Kelowna, BC, V1V 1V7 Canada (email:{jahangir.hossain, julian.cheng}@ubc.ca).

been made on the first topic, the scenarios under which are point-to-point systems [14], [18], single-cell systems [15], [19], [26], [27] and multi-cell systems [16], [17], [20]–[22], [25]. For example, considering the non-causal channel state information (CSI) and non-causal energy state information (ESI), the authors in [14] optimized the total energy cost of the point-to-point system under the energy harvesting constraint and the outage constraint over multiple fading slots. In [19], several online algorithms were proposed to minimize the grid energy expenditure based on a prior knowledge of hourly-varying energy price. The authors in [21] investigated both multi-stage energy allocation problem and multi-cell energy balancing problem, and proposed power allocation and energy balancing algorithms to minimize the grid energy expenditure. For the second topic, the authors in [11] studied the power allocation problem in the point-to-point HPC systems, and both online and offline algorithms were proposed to maximize the throughput. In [23], they further investigated the energy efficiency in the orthogonal frequency division multiple access (OFDMA) based HPC systems. The authors in [24] proposed the EH based cognitive radio systems, where the secondary users are solely powered by the renewable energy. The tradeoff between the expected throughput of the EH secondary user and spectrum sensing was investigated. Besides, the authors in [18] investigates both the data rate maximization problem and the grid energy minimization problem in hybrid powered point-to-point systems via offline optimization and online optimization. Although the works in [11], [14]–[24] consider uncertainty of the CSI and the ESI, it is difficult to implement the proposed algorithms in practical systems because the statistical properties of the CSI and the ESI are either difficult or costly to obtain. Therefore, it is preferable to design online algorithms that are independent of the statistical distributions of the CSI and the ESI.

Limited literature investigated the beamforming design in the HPC systems [17], [20], [22], [25]–[27]. Using the bounded CSI model, robust beamforming algorithms were proposed to minimize the short-term worst-case grid expenditure [20] and the long-term grid expenditure [22] subject to the worst-case SINR requirement for the coordinated multipoint systems. However, the harvested renewable energy is stored in batteries for further usage, which still suffers from the imperfection of the batteries. The authors in [17], [25] first proposed the HUT strategy in the HPC systems. Yet, the algorithms in [17], [25] are based on the instantaneous CSI and ESI, and the accumulated impact of previous CSI and ESI are ignored, which cannot be used to optimize the long-term performance metrics. Exploiting the uplink-downlink duality of capacity region of dirty paper code, the authors investigated the power allocation over the finite scheduling horizon of the HPC systems in [26]. Later, the authors in [27] extended the scheduling horizon to infinity, and solved the power allocation in the HPC systems via the stochastic subgradient method. Different from [17], [20], [22], [25], we study the long-term grid energy expenditure minimization subject to the packet reception rate requirements, which is modeled as sigmoid function. Instead using the well-known log-concave function for the data rate [15], [17], [20], [22]–[27], we model the packet reception rate as a sigmoid function, which captures effects of packet transmission failure and the data rate. Therefore, algorithm design based on the sigmoidal packet rate provides a realistic insight of the practical systems. However, the non-convexity of the sigmoid function gives new challenges in the beamforming algorithm design.

### B. Contributions

We study the long-term grid energy expenditure minimization problem in the HPC systems with HUT strategy. In our system model, the BS is equipped with both smart meter and energy harvesters such that the BS is able to perform the two-way energy trading. Different from previous works in [15], [17], [20], [22]–[27], we consider the joint effect of the data rate and the transmission reliability during the packet transmission. The contributions of this work are summarized as follows.

- By taking into account of the uncertainties of the renewable energy arrival and the packet arrival, we formulate the long-term grid energy expenditure minimization problem via the stochastic optimization theory [28]. In the formulated problem, we consider the effects of the channel quality and the packet reception failure while designing beamforming algorithms. Therefore, the considered optimization problem is a cross-layer one.
- By leveraging the Lyapunov optimization technique, we reformulate the stochastic optimization problem, which is named as a per-frame grid energy plus weighted packet rate (GEWPR) minimization problem. The problem is NP-hard.
- Two suboptimal beamforming algorithms are proposed based on the successive approximation beamforming (SABF) technique and the zero-forcing beamforming (ZFBF) technique. The convergence properties are established, and the corresponding computational complexities are analyzed.

Numerical results are used to verify the performance of the packet delay and the grid energy expenditure of the proposed algorithms. By tuning a control parameter $V$, the suboptimal grid energy expenditure can be traded for the packet delay of each wireless node (WN). The proposed iterative SABF algorithm can achieve a lower packet delay and a lower grid energy expenditure than that of the ZFBF algorithm.

The rest of this paper is organized as follows. In Section II, we describe the HPC system model and formulate the optimization problem. The tradeoff between the grid energy expenditure and the packet delay is revealed via the stochastic optimization technique, and a dynamic beamforming framework is proposed in Section III. In Section IV, two different techniques are used to facilitate the development of low-complexity beamforming algorithms in each frame. Simulation results are provided in Section V, and Section VI concludes this paper.

*Notations:* Boldface letters refer to vectors (lowercase) or matrices (uppercase). $\mathbb{C}$ and $\mathbb{N}_+$ denote the set of complex numbers and the set of positive integers, respectively. $\|\cdot\|_2$ refers to $\ell_2$-norm. $(\cdot)^{\mathrm{T}}$, $(\cdot)^{\mathrm{H}}$ and $(\cdot)^{\dagger}$ denote, respectively, the transpose operation, the conjugate transpose operation and the pseudo-inverse operation. The notation $\text{diag}(\boldsymbol{w})$ stands for the diagonal matrix whose diagonal elements consist of $\boldsymbol{w}$. $\boldsymbol{I}$ and $\boldsymbol{0}$ denote an identity matrix and an all-zero matrix, respectively.

## II. SYSTEM MODEL AND PROBLEM FORMULATION

### A. Overall System Description

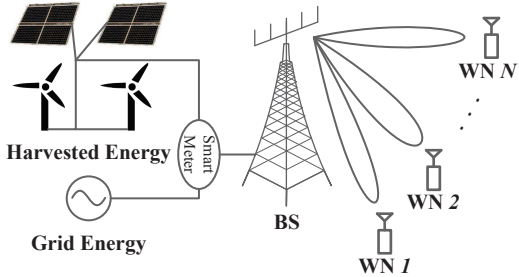

Fig. 1. An illustration of a single cell HPC system, where the BS can use energy from both smart grid and the harvested energy via solar panels and wind turbines.

We consider the downlink transmission of an HPC system that consists of a BS and $N$ WNs. The BS is equipped with $N_T > 1$ antennas, and each WN is equipped with a single antenna. Without loss of generality, we assume there is no correlation among the transmission antennas at the BS[1]. As shown in Fig. 1, the BS can use the energy from both smart grid and energy harvesting devices, such as solar panels and wind turbines. The BS can obtain the instantaneous CSI by exploiting the channel reciprocity and hand shaking signals. Let $\mathcal{N} = \{1, 2, \ldots, N\}$ denote the set of WNs in the HPC system. All the WNs receive information from the BS via beamforming in the downlink period. The HPC system operates in discrete-time mode with the index of $t$-th frame denoting a unit duration $[t, t+1)$, $t = \{0, 1, 2, \ldots\}$; therefore, the "power" and "energy" can be used interchangeably. We assume the channel fading and renewable energy arrival follow the block-based model, where the channel coefficients and the EH rate remain constant during each frame and vary over different time scales. Specifically, the channel coefficients change over different frames, and the EH rate fluctuates over several frames. This assumption is reasonable because the coherence time of the renewable energy arrival is generally longer than the channel coherence time [31]. In this paper, we do not consider the energy storage capability at the BS due to the low utilization efficiency of the renewable energy. Via the smart meter, the BS can either purchase energy from the smart grid due to the shortage of the harvested energy or sell the extra harvested energy to the smart grid in order to reduce the grid energy expenditure of the HPC system.

[1]This corresponds to antennas that are physically separated by at least half of the wavelength [29], [30].

### B. Signal Model

In this paper, each BS-WN link is assigned with a beamforming vector. Hence, the transmitted signal in the $t$-th frame at the BS is denoted by

$$x_{\text{Tx}}(t) = \sum_{n=1}^{N} \boldsymbol{w}_n(t) x_n(t) \quad (1)$$

where $\boldsymbol{w}_n(t) \in \mathbb{C}^{N_T \times 1}$ and $x_n(t)$, respectively, represent the beamforming vector and the information-bearing signal for the $n$-th BS-WN link in the $t$-th frame, where $\mathbb{E}[|x_n(t)|^2] = 1$ with $\mathbb{E}[\cdot]$ denoting the expectation operator. Thus, the received signal at the $n$-th WN in the $t$-th frame is denoted by

$$y_n(t) = \sum_{m=1}^{N} \boldsymbol{h}_n^{\text{H}}(t) \boldsymbol{w}_m(t) x_m(t) + z_n(t), n \in \mathcal{N} \quad (2)$$

where $z_n(t) \sim \mathcal{CN}(0, \sigma_n^2)$ denotes the noise-plus-background-interference[2] at the $n$-th WN; $\boldsymbol{h}_n(t) \in \mathbb{C}^{N_T \times 1}$ is the complex channel coefficient vector of the $n$-th BS-WN link in the $t$-th frame; each entry of $\boldsymbol{h}_n(t)$ is a circularly-symmetric complex Gaussian random variable with mean zero and variance $d_n^{-\chi}$, where $d_n$ and $\chi$ are, respectively, the link length of the $n$-th BS-WN link and the pathloss exponent. Here, we note that the vector $\boldsymbol{h}_n(t)$ captures the composite effects of multipath fading and pathloss. In (2), $\boldsymbol{w}_m(t)$ and $x_m(t)$ represent the beamforming vector and the information-bearing signal of the $m$-th BS-WN link in the $t$-th frame, respectively.

The received signal-to-interference-plus-noise ratio (SINR) at the $n$-th WN in the $t$-th frame is denoted as

$$\text{SINR}_n(t) = \frac{\left|\boldsymbol{h}_n^{\text{H}}(t) \boldsymbol{w}_n(t)\right|^2}{\sum_{m=1, m \neq n}^{N} \left|\boldsymbol{h}_n^{\text{H}}(t) \boldsymbol{w}_m(t)\right|^2 + \sigma_n^2}. \quad (3)$$

On the other hand, a certain amount of power is consumed to enable the information transmission [3]. In particular, the power consumption of the considered HPC system in the $t$-th frame, $P_{\text{tot}}(t)$, is denoted as

$$P_{\text{tot}}(t) = \frac{1}{\psi} \sum_{n=1}^{N} \|\boldsymbol{w}_n(t)\|_2^2 + P_{\text{sp}} \quad (4)$$

where the constant $\psi$ is the power amplifier efficiency; the term $P_{\text{sp}}$ is the power consumption of signal processing, and it is calculated as [3]

$$P_{\text{sp}} = P_{\text{sp,b}} \left(0.87 + 0.1 N_T + 0.03 N_T^2\right) \quad (5)$$

where $P_{\text{sp,b}}$ is set at 115mW in the HPC system. In (5), the linear term of $N_T$ represents energy overhead due to multiple-input single-output (MISO) pilots, and the quadratic term stands for MISO signal processing overhead.

[2]We model the noise-plus-background-interference as a circularly symmetric complex Gaussian random variable due to the large quantities of interferers from the other cells [32].

## C. Grid Energy Cost Model

Assuming the rate of harvesting energy[3] remains constant during each frame, we obtain the amount of harvested energy by the BS during $t$-th frame as $E_{\text{hav}}(t)$. We set the price for purchasing (selling) a unit power from (to) the smart grid as $a_b > 0$ ($a_s > 0$). Thus, the grid energy expenditure in the $t$-th frame is calculated as

$$G(t) = a_b[P_{\text{tot}}(t) - E_{\text{hav}}(t)]^+ - a_s[E_{\text{hav}}(t) - P_{\text{tot}}(t)]^+ \quad (6)$$

where $[x]^+ = \max[x, 0]$. To avoid the operator of the HPC system making non-justifiable profit, the operator of the smart grid should set $a_b \geq a_s > 0$.

## D. Packet Correct Reception Rate and Traffic Queues

For the wireless data traffic, multiple modulation and coding schemes (MCSs) can be used to achieve tradeoff between the data rate and the transmission reliability [30]. High order modulation schemes, which allow more information bits to be transmitted per symbol, shorten the Euclidean distance of the signal constellation points. Therefore, more errors occur in the decoding process. Various coding schemes, accompanied with the modulation schemes, are used to adapt to the channel variations. Decreasing the coding rate leads to a decrease in the effective data rate.

Generally, a packet consists of $L$ information bits. An error occurs when one of the $L$ information bits is erroneously decoded. Hence, the correct packet reception rate at the $n$-th WN can be denoted by

$$\Pi_n(t) = M_n \prod_{l=1}^{L} (1 - \pi_l(\text{SINR}_n(t))) \quad (7)$$

where $M_n$ and $\pi_l(\text{SINR}_n(t))$ denote, respectively, the packet transmission rate for the $n$-th WN and the error probability of the $l$-th information bits of the packet in the $t$-th frame. Using the approximation method developed in [33], we obtain the correct packet reception rate as

$$\Pi_n(t) = \frac{M_n}{1 + \exp(-c_n(10\log_{10}(\text{SINR}_n(t)) - b_n))} \quad (8)$$

where $c_n$ and $b_n$ are MCS specific parameters. Dividing the packet departure process and packet arrival process at the $n$-th WN by $M_n$, the normalized packet departure process of the $n$-th WN in the BS is denoted by

$$U_n(t) = \frac{1}{1 + \exp(-c_n(10\log_{10}(\text{SINR}_n(t)) - b_n))}. \quad (9)$$

As the BS, $N$ traffic queues are used to buffer the random arrival packets for the corresponding $N$ BS-WN links in the HPC system. We denote the normalized backlog[4] of the $n$-th traffic queue in the $t$-th frame as $q_n(t)$. As a result, the backlog of the $n$-th traffic queue evolves as

$$q_n(t+1) = [q_n(t) - U_n(t)]^+ + U_{n,\text{req}}(t), n \in \mathcal{N} \quad (10)$$

where $U_{n,\text{req}}(t)$ denotes the normalized packet arrival process with $U_{n,\text{req}}(t) \in [0, 1]$. Here, the normalized packet arrival process $U_{n,\text{req}}(t)$ is independent and identically distributed over all frames with $\mathbb{E}[U_{n,\text{req}}(t)] = u_{n,\text{req}}$.

*Definition 1:* (**Mean Rate Stable** [28]) A discrete-time queue $\Xi(t)$ is mean rate stable if $\lim_{t \to \infty} \frac{\mathbb{E}[|\Xi(t)|]}{t} = 0$.

A mean rate stable discrete-time queue indicates that the time average arrival rate is smaller than or equal to the time average departure rate.

## E. Problem Formulation

Based on the aforementioned description, the grid energy expenditure minimization problem in the HPC with HUT is formulated mathematically. After obtaining the instantaneous CSI and the ESI, the BS performs the dynamic beamforming with the following constraints:

- Traffic queue constraints:

    Traffic queues $q_n(t)$ are mean rate stable, $n \in \mathcal{N}$. (11)

- WNs' SINR constraints:

$$\text{SINR}_n(t) \geq \Gamma_{n,\text{req}}, n \in \mathcal{N} \quad (12)$$

    where $\Gamma_{n,\text{req}}$ denotes the instantaneous SINR requirements of the $n$-th WN. This set of constraints is used to guarantee that the packet error probability of each WN is at an acceptable level in the $t$-th frame.

- Transmission power constraint: Due to the circuit limitation, the transmission power is limited under a threshold

$$\sum_{n=1}^{N} \|\boldsymbol{w}_n(t)\|_2^2 \leq P^{\max} \quad (13)$$

    where $P^{\max}$ represents the maximum transmission power of the BS.

In this paper, we perform the dynamic beamforming to minimize the long-term grid energy expenditure in the HPC system with HUT strategy under the traffic queue constraints, WNs' SINR constraints and transmission power constraint. To incorporate the uncertainty of the renewable energy arrival, the objective function is defined as the time average expectation of the grid energy expenditure. Therefore, we can formulate the corresponding problem via a stochastic optimization model as

$$\min_{\boldsymbol{w}_n(t)} \lim_{T \to \infty} \frac{1}{T} \sum_{t=0}^{T-1} \mathbb{E}[G(t)] \quad (14a)$$

$$\text{s.t. } (11) - (13). \quad (14b)$$

The optimization problem in (14) is cross-layer optimization because it considers the effects of the channel quality and the packet reception failure while designing beamforming algorithms. Moreover, the optimization problem in (14) can also be used to reveal the tradeoff between the grid energy expenditure and packet delay quantitatively.

---

[3] The BS can obtain real-time solar radiation rate and wind rate. Based on the real-time solar radiation rate and the wind rate, the BS can estimate the amount of harvested energy due to the fact that the real-time solar radiation rate and the wind rate change slowly when compared with the variation of the channel state information.

[4] We note that the packet departure process and the packet arrival process are normalized by $M_n$. Hence, the backlog of the traffic queue is also normalized by $M_n$.

*Remark 1:* In our formulation, we consider that HPC system utilizes the HUT strategy to use the harvested energy; therefore, we only consider the traffic queues of the UEs. The case that jointly consider the traffic queues and the energy queue will be investigated in our future work.

## III. DYNAMIC CROSS-LAYER BEAMFORMING VIA STOCHASTIC OPTIMIZATION

By leveraging the stochastic optimization theory [28], we reformulate (14) as a problem whose objective function is the grid energy expenditure plus the weighted packet rate. We deal with the time average issue in the objective function and the constraints of the optimization problem (14) by constructing a Lyapunov drift-plus-penalty function with respect to (14a) and (11) as

$$\Delta(\boldsymbol{q}(t)) + V\mathbb{E}[G(t)|\boldsymbol{q}(t)] \quad (15)$$

where $\boldsymbol{q}(t) \triangleq [q_1(t), q_2(t), \ldots, q_N(t)]$ and the introduced control parameter $V > 0$; the one frame conditional Lyapunov drift function $\Delta(\boldsymbol{q}(t))$ is defined as

$$\Delta(\boldsymbol{q}(t)) \triangleq \frac{1}{2}\mathbb{E}\left[\|\boldsymbol{q}(t+1)\|_2^2 \,\Big|\, \boldsymbol{q}(t)\right] - \frac{1}{2}\|\boldsymbol{q}(t)\|_2^2 \quad (16)$$

where $\mathbb{E}[\cdot]$ is for the channel coefficients $\{\boldsymbol{h}_n(t)\}_{n\in\mathcal{N}}$ and the beamforming vectors $\boldsymbol{w}_n(t)$.

*Lemma 1:* Supposing the channel coefficients $\{\boldsymbol{h}_n(t)\}_{n\in\mathcal{N}}$, are independent and identically distributed over different frames. We obtain the upper bound of the Lyapunov drift-plus-penalty in (15) as [28]

$$\Delta(\boldsymbol{q}(t)) + V\mathbb{E}[G(t)|\boldsymbol{q}(t)] \leq N + V\mathbb{E}[G(t)|\boldsymbol{q}(t)]$$
$$+ \sum_{n=1}^{N} q_n(t)\mathbb{E}[U_{n,\text{req}}(t) - U_n(t)|\boldsymbol{q}(t)]. \quad (17)$$

*Proof:* See Appendix A. □

It can be shown that any solution that minimizes right-hand side (RHS) of (17) subject to (12) and (13) given the channel coefficients $\{\boldsymbol{h}_n(t)\}_{n\in\mathcal{N}}$ and the backlog of traffic queues $\boldsymbol{q}(t)$ is also a solution to the original problem (14). The detailed proof will be presented in Proposition 1. Thus, the RHS of (17) is useful in developing the dynamic cross-layer beamforming algorithm. The introduced control parameter $V$ is used to control the tradeoff between the grid energy expenditure and the average backlog of the traffic queues.

In the RHS of (17), we observe that the constant $N$ and the term $\mathbb{E}[U_{n,\text{req}}(t)|\boldsymbol{q}(t)]$ have no effect on designing the beamforming vectors $\boldsymbol{w}_n(t)$. Besides, minimizing the conditional expectation terms $\mathbb{E}[G(t)|\boldsymbol{q}(t)]$ and $\mathbb{E}[-U_n(t)|\boldsymbol{q}(t)]$ can be manipulated as minimizing $G(t)$ and $-U_n(t)$ given $\boldsymbol{q}(t)$ and $\{\boldsymbol{h}_n(t)\}_{n\in\mathcal{N}}$ according to the principle of opportunistically minimizing an expectation [28, Section 1.8]. As a result, we can simplify the original optimization problem (14) according to (17) as the GEWPR minimization problem

$$\min_{\boldsymbol{w}_n(t)} VG(t) - \sum_{n=1}^{N} q_n(t)U_n(t) \quad (18a)$$

$$\text{s.t.} \sum_{n=1}^{N} \|\boldsymbol{w}_n(t)\|_2^2 \leq P^{\max} \quad (18b)$$

$$\text{SINR}_n(t) \geq \Gamma_{n,\text{req}}, \forall n. \quad (18c)$$

The optimization problem (18) has non-convex objective function and non-convex constraints. Following the similar arguments in [34], we can prove the optimization problem (18) is NP-hard. Hence, we are motivated to obtain the suboptimal solutions to the optimization problem (18). Based on the beamforming technique, the methods to check the feasibility are different. We will review the feasibility check methods separately for the beamforming techniques.

---

**Algorithm 1** Dynamic Beamforming Algorithm
---
1: At the start of each frame, the BS observes the backlogs of traffic queues $q_n(t)$, complex channel coefficient vectors $\{\boldsymbol{h}_n(t)\}_{n\in\mathcal{N}}$ and the amount of harvested energy $E_{\text{hav}}(t)$.
2: **if** the optimization problem (18) is feasible **then**
3:     The BS obtains the utility of each WN via solving the optimization problem (18) suboptimally.
4: **else**
5:     The BS obtains the utility of each WN via randomly choosing the beamforming vector $\{\boldsymbol{w}_n(t)\}_{n\in\mathcal{N}}$, under the transmission power constraint (18b).
6: **end if**
7: The BS updates the backlogs of virtual queues $q_n(t)$ according to (18).

---

*Proposition 1:* With the independent and identically distributed $\{\boldsymbol{h}_n(t)\}_{n\in\mathcal{N}}$, and supposing $[u_{1,\text{req}}, \ldots, u_{N,\text{req}}] \in \boldsymbol{\lambda} + \epsilon$, where $\boldsymbol{\lambda}$ and $\epsilon$ denotes the capacity region of (14) achieved by the suboptimal solutions and a small positive constant. The initial expected backlog of traffic queues satisfies $\mathbb{E}\left[\|\boldsymbol{q}(t)\|_2^2\right] < \infty$. The properties of Algorithm 1 are described as follows:

1) Each virtual queue $q_n(t)$ is mean rate stable, $n \in \mathcal{N}$.
2) The total grid energy expenditure is upper bounded by

$$\lim_{t\to\infty} \frac{1}{t} \sum_{\tau=0}^{t-1} \mathbb{E}[G(\tau)] \leq \frac{N}{V} + G^{\text{SUBOPT}} \quad (19)$$

where $G^{\text{SUBOPT}}$ is the maximum suboptimal value of the grid energy cost in (18).

3) The average queue backlog of the traffic queues is upper bounded by

$$\lim_{t\to\infty} \frac{1}{t} \sum_{\tau=0}^{t-1} \sum_{n=1}^{N} \mathbb{E}[q_n(\tau)] \leq \frac{N + V\left(G^{\text{SUBOPT}} - G^{\min}\right)}{\epsilon} \quad (20)$$

where $G^{\min}$ is the smallest value of grid energy expenditure.

*Proof:* See Appendix B. □

We obtain two conclusions from *Proposition 1*: 1) the output of the Algorithm 1 is a feasible solution to the optimization problem in (14); and 2) the tradeoff between the grid energy expenditure and the packet delay is quantitatively revealed.

*Remark 2:* The proposed dynamic cross-layer beamforming framework does not have limitation on the number of antennas of the WNs. Hence, the developed dynamic beamforming framework can be used to the HPC systems with multiple-input and multiple-output channels.

## IV. SUBOPTIMAL SOLUTIONS

### A. Beamforming for a Given Frame

For brevity, we omit the frame index in this section. We note that the constraints in (18c) require the BS to consume a certain amount of power. However, the transmission power of the BS is limited. The optimization problem (18) can be infeasible under certain CSI and network scenarios when $\Gamma_{n,\mathrm{req}}$ is too high, $n \in \mathcal{N}$. Before solving (18), the feasibility of (18) should be checked via solving the following optimization problem

$$\text{find } \{\boldsymbol{w}_n\}$$
$$\text{s.t. } \sum_{n=1}^{N} \|\boldsymbol{w}_n\|_2^2 \le P^{\max} \tag{21}$$
$$\text{SINR}_n \ge \Gamma_{n,\mathrm{req}}, \forall n.$$

The optimization problem (21) can be solved by the standard optimization techniques, such as a second-order cone programming (SOCP) [35] and uplink-downlink duality based algorithm [36].

To deal with the non-convexity of the $\text{SINR}_n$ in the objective function (18a), we introduce positive auxiliary variables $\{\alpha_n\}_{n \in \mathcal{N}}$. After some mathematical manipulation, the optimization problem (18) is recast as

$$\min_{\boldsymbol{w}_n, \alpha_n} VG - \sum_{n=1}^{N} q_n \frac{1}{1 + \exp\left(-c_n \left(10 \log_{10}(\alpha_n) - b_n\right)\right)} \tag{22a}$$
$$\text{s.t. } \sum_{n=1}^{N} \|\boldsymbol{w}_n\|_2^2 \le P^{\max} \tag{22b}$$
$$\text{SINR}_n \ge \Gamma_{n,\mathrm{req}}, \forall n \tag{22c}$$
$$\text{SINR}_n \ge \alpha_n, \forall n. \tag{22d}$$

Note that the constraints in (22d) are active at each suboptimal point; otherwise, the objective function (22a) can take a strictly smaller value by increasing $\{\alpha_n\}_{n \in \mathcal{N}}$. Based on this observation, we can also conclude that each local optimal value of $\alpha_n$ makes the following inequality holds, i.e.,

$$\alpha_n^* \ge \Gamma_{n,\mathrm{req}}, \forall n \tag{23}$$

where $\alpha_n^*$ represents a local optimal value of $\alpha_n$.

The major challenges in solving (22) are two folds: 1) the sum-of-ratios component in the objective function (22a); 2) the non-convex constraints (22c) and (22d).

We leverage the Lagrangian duality theorem to deal with sum-of-ratios component in (22a) and obtain an equivalent form of the optimization problem (22).

*Proposition 2:* If $\{\boldsymbol{w}_n, \alpha_n\}_{n \in \mathcal{N}}$ satisfy the Karush-Kuhn-Tucker (KKT) conditions of the optimization problem (22), we can conclude that there exist parameters

$$\gamma_n = \frac{1}{1 + \exp\left(-c_n \left(10 \log_{10}(\alpha_n) - b_n\right)\right)} \tag{24a}$$
$$\varpi_n = \frac{q_n}{1 + \exp\left(-c_n \left(10 \log_{10}(\alpha_n) - b_n\right)\right)} \tag{24b}$$

such that an optimization problem has the same KKT conditions as in (22), which is shown as

$$\min_{\boldsymbol{w}_n, \alpha_n} VG + \sum_{n=1}^{N} \varpi_n \gamma_n \exp\left(-c_n \left(10 \log_{10}(\alpha_n) - b_n\right)\right) \tag{25a}$$
$$\text{s.t. } \sum_{n=1}^{N} \|\boldsymbol{w}_n\|_2^2 \le P^{\max} \tag{25b}$$
$$\text{SINR}_n \ge \Gamma_{n,\mathrm{req}}, \forall n \tag{25c}$$
$$\text{SINR}_n \ge \alpha_n, \forall n. \tag{25d}$$

*Proof:* See Appendix C. □

We note that the grid energy expenditure $G$ in (25a) can be reformulated as

$$G = (a_b - a_s)[P_{\mathrm{tot}} - E_{\mathrm{hav}}]^+ + a_s[P_{\mathrm{tot}} - E_{\mathrm{hav}}] \tag{26}$$

which is a convex function. In (25a), the second term $\sum_{n=1}^{N} \varpi_n \gamma_n \exp\left(-c_n \left(10 \log_{10}(\alpha_n) - b_n\right)\right)$ is strictly convex. Hence, the objective function in (25a) is a strictly convex function.

Proposition 2 shows that the objective function in sum-of-ratios form of the optimization problem (22) can be equivalently transformed into a summation form with introduced auxiliary variables. However, the objective function in the summation form is strictly convex, which reduces the complexity in the design of the algorithm. Hereinafter, the design of the algorithm is based on the optimization problem (25) with (24a) and (24b).

### B. Convex Approximation Based Beamforming

To deal with the nonconvexity, we use the successive approximation technique. Introducing another set of auxiliary variables $\beta_n$, $n \in \mathcal{N}$, we can reformulate the optimization problem (25) as

$$\min_{\boldsymbol{w}_n, \alpha_n, \beta_n} \sum_{n=1}^{N} \varpi_n \gamma_n \exp\left(-c_n \left(10 \log_{10}(\alpha_n) - b_n\right)\right)$$
$$+ VG \tag{27a}$$
$$\text{s.t. } \sum_{n=1}^{N} \|\boldsymbol{w}_n\|_2^2 \le P^{\max} \tag{27b}$$
$$\boldsymbol{h}_n^{\mathrm{H}} \boldsymbol{w}_n \ge \beta_n \sqrt{\Gamma_{n,\mathrm{req}}}, \forall n \tag{27c}$$
$$\boldsymbol{h}_n^{\mathrm{H}} \boldsymbol{w}_n \ge \beta_n \sqrt{\alpha_n}, \forall n \tag{27d}$$
$$\sqrt{\sigma_n^2 + \sum_{m=1, m \ne n}^{N} \left|\boldsymbol{h}_n^{\mathrm{H}} \boldsymbol{w}_m\right|^2} \le \beta_n, \forall n \tag{27e}$$
$$\mathrm{Im}\left(\boldsymbol{h}_n^{\mathrm{H}} \boldsymbol{w}_n\right) = 0, \forall n. \tag{27f}$$

We justify the equivalence between the optimization problems (25) and (27) via the following two perspectives: 1) forcing the phase of the term $\boldsymbol{h}_n^{\mathrm{H}}\boldsymbol{w}_n$ to zero does not change the value of the objective function as rotating the phase of $\boldsymbol{w}_n$ leads to the same value of $\left|\boldsymbol{h}_n^{\mathrm{H}}\boldsymbol{w}_n\right|$; 2) constraints in (27e) are active, and see Appendix D for detailed proof.

We observe that the objective function in the optimization problem (27) is convex, and the only set of non-convex constraints is in (27d). The constraints in (27d) are equivalent to

$$\frac{\left|\boldsymbol{h}_n^{\mathrm{H}}\boldsymbol{w}_n\right|^2}{\alpha_n} \geq \beta_n^2, \forall n \quad (28)$$

where the left hand side is a quadratic-over-linear term, which is a joint convex function of $\boldsymbol{w}_n$ and $\alpha_n$ and has a linear lower bound as

$$\frac{\left|\boldsymbol{h}_n^{\mathrm{H}}\boldsymbol{w}_n\right|^2}{\alpha_n} \geq \frac{2\mathrm{Re}\left(\left(\boldsymbol{w}_n^{(\tau)}\right)^{\mathrm{H}}\boldsymbol{h}_n\boldsymbol{h}_n^{\mathrm{H}}\boldsymbol{w}_n\right)}{\alpha_n^{(\tau)}} - \left(\frac{\left|\boldsymbol{h}_n^{\mathrm{H}}\boldsymbol{w}_n^{(\tau)}\right|}{\alpha_n^{(\tau)}}\right)^2 \alpha_n$$
$$= \Psi_n^{(\tau)}(\boldsymbol{w}_n, \alpha_n), \forall n \quad (29)$$

where $\tau$ represents the iteration index; $\boldsymbol{w}_n^{(\tau)}$ and $\alpha_n^{(\tau)}$ are, respectively, the beamforming vector and the auxiliary variable for the $n$-th WN-BS link in the $\tau$-th iteration. Using (29), we obtain the convex approximation of the non-convex constraints in (27d) successively in the $\tau$-th iteration as

$$\Psi_n^{(\tau)}(\boldsymbol{w}_n, \alpha_n) \geq \beta_n^2. \quad (30)$$

As a result, we finally obtain the following convex optimization problem as

$$\min_{\boldsymbol{w}_n, \alpha_n, \beta_n} VG + \sum_{n=1}^{N} \varpi_n \gamma_n \exp\left(-c_n\left(10\log_{10}(\alpha_n) - b_n\right)\right)$$
$$\text{s.t.} \sum_{n=1}^{N} \|\boldsymbol{w}_n\|_2^2 \leq P^{\max}$$
$$\boldsymbol{h}_n^{\mathrm{H}}\boldsymbol{w}_n \geq \beta_n\sqrt{\Gamma_{n,\mathrm{req}}}, \forall n$$
$$\Psi_n^{(\tau)}(\boldsymbol{w}_n, \alpha_n) \geq \beta_n^2, \forall n$$
$$\left\|\left[\sigma_n, \boldsymbol{h}_n^{\mathrm{H}}\boldsymbol{W}_{-n}\right]\right\|_2 \leq \beta_n, \forall n$$
$$\mathrm{Im}\left(\boldsymbol{h}_n^{\mathrm{H}}\boldsymbol{w}_n\right) = 0, \forall n$$
$$(31)$$

where $\boldsymbol{W}_{-n} = [\boldsymbol{w}_1, \ldots, \boldsymbol{w}_{n-1}, \boldsymbol{w}_{n+1}, \ldots, \boldsymbol{w}_N], n \in \mathcal{N}$. Based on (24a) and (24b), we summarize the procedure of the SABF algorithm in Algorithm 2.

***Proposition 3:*** The proposed SABF algorithm converges to a point that satisfies the KKT conditions of (22).

*Proof:* See Appendix E. □

## C. Zero-Forcing Beamforming

In previous subsection, we have obtained the SABF algorithm for the optimization problem (22) which introduces multiple auxiliary variables. As a result, solving such a problem incurs high computation complexity. Therefore, we consider in this subsection a suboptimal solution with lower computation complexity via the ZFBF technique, where the beamforming vectors are designed to null the interference among WNs. However, to perform the ZFBF, the number of transmission antennas should be equal to or larger than the number of WNs, i.e., $N_T \geq N$.

**Algorithm 2** SABF Algorithm
1: BS sets the iteration index $\tau = 0$, the stop threshold $\epsilon$, the maximum number of iteration $T^{\max}$ and initial traffic queue backlog $q_n(t)$.
2: BS obtains a feasible point $\left\{\boldsymbol{w}_n^{(0)}\right\}_{n \in \mathcal{N}}$.
3: BS updates $\left\{\varpi_n^{(0)}\right\}_{n \in \mathcal{N}}$, $\left\{\gamma_n^{(0)}\right\}_{n \in \mathcal{N}}$ and $\left\{\alpha_n^{(0)}\right\}_{n \in \mathcal{N}}$ via (70), (71) and (3), respectively.
4: **repeat**
5: $\quad \tau := \tau + 1$
6: $\quad$ With $\left\{\boldsymbol{w}_n^{(\tau-1)}, \gamma_n^{(\tau-1)}, \varpi_n^{(\tau-1)}\right\}_{n \in \mathcal{N}}$, BS solves the optimization problem (31) and obtains the beamforming vector $\boldsymbol{w}_n^{(\tau)}$ and corresponding auxiliary variables $\alpha_n^{(\tau)}$ and $\beta_n^{(\tau)}$ for $\tau$-th iteration.
7: $\quad$ BS updates the following parameters:
$$\boldsymbol{w}_n^{(\tau)} = \boldsymbol{w}_n^{(\tau-1)} \quad (32)$$
$$\gamma_n^{(\tau)} = \frac{1}{1 + \exp\left(-c_n\left(10\log_{10}\left(\alpha_n^{(\tau-1)}\right) - b_n\right)\right)} \quad (33)$$
$$\varpi_n^{(\tau)} = \frac{q_n}{1 + \exp\left(-c_n\left(10\log_{10}\left(\alpha_n^{(\tau-1)}\right) - b_n\right)\right)} \quad (34)$$
where $\boldsymbol{\gamma}^{(\tau)} \triangleq \left[\gamma_1^{(\tau)}, \gamma_2^{(\tau)}, \ldots, \gamma_n^{(\tau)}\right]$ and $\boldsymbol{\varpi}^{(\tau)} \triangleq \left[\varpi_1^{(\tau)}, \varpi_2^{(\tau)}, \ldots, \varpi_n^{(\tau)}\right]$.
8: **until** $\frac{\left\|\boldsymbol{\gamma}^{(\tau)} - \boldsymbol{\gamma}^{(\tau-1)}\right\|_2}{\left\|\boldsymbol{\gamma}^{(\tau-1)}\right\|_2} \leq \epsilon$ and $\frac{\left\|\boldsymbol{\varpi}^{(\tau)} - \boldsymbol{\varpi}^{(\tau-1)}\right\|_2}{\left\|\boldsymbol{\varpi}^{(\tau-1)}\right\|_2} \leq \epsilon$ or $\tau > T^{\max}$.

To derive the ZFBF vector, we first decouple the power $p_n$ of the $n$-th WN from the corresponding beamforming vector $\boldsymbol{w}_n, n \in \mathcal{N}$. Then, we define the complex channel coefficient matrix and the beamforming matrix for all WNs, respectively, as $\boldsymbol{H} \triangleq [\boldsymbol{h}_1, \boldsymbol{h}_2, \ldots, \boldsymbol{h}_n]$ and $\boldsymbol{W} \triangleq [\boldsymbol{w}_1, \boldsymbol{w}_2, \ldots, \boldsymbol{w}_n]$. Accordingly, the signal model (2) can be shown in compact form as

$$\boldsymbol{y} = \boldsymbol{H}^{\mathrm{H}}\boldsymbol{W}\,\mathrm{diag}\left(\sqrt{\boldsymbol{p}}\right)\boldsymbol{x} + \boldsymbol{z} \quad (35)$$

where the power vector, signal vector and noise vector are respectively denoted by $\boldsymbol{p} \triangleq [p_1, p_2, \ldots, p_n]$, $\boldsymbol{x} \triangleq [x_1, x_2, \ldots, x_n]$ and $\boldsymbol{z} \triangleq [z_1, z_2, \ldots, z_n]$.

Thus, the ZFBF matrix is written as

$$\boldsymbol{W}_{\mathrm{ZF}} = \left(\boldsymbol{H}^{\mathrm{H}}\right)^{\dagger} = \boldsymbol{H}\left(\boldsymbol{H}^{\mathrm{H}}\boldsymbol{H}\right)^{-1}. \quad (36)$$

With (36), we obtain the beamforming vector for the $n$-th BS-WN link under the ZFBF technique as

$$\boldsymbol{w}_{n,\mathrm{ZF}} = \boldsymbol{W}_{\mathrm{ZF}}(:, n), \forall n. \quad (37)$$

Because $\boldsymbol{H}^{\mathrm{H}}\boldsymbol{W}_{\mathrm{ZF}} = \boldsymbol{I}$, we obtain the received signal model for all WNs as $\boldsymbol{y} = \mathrm{diag}\left(\sqrt{\boldsymbol{p}}\right)\boldsymbol{x} + \boldsymbol{z}$. As a result, the received signal-to-noise ratio (SNR) of the $n$-th WN is denoted by

$$\mathrm{SNR}_n = \frac{p_n}{\sigma_n^2}, \forall n. \quad (38)$$

Using (37) and (38), we can obtain the optimization problem (25) under the ZFBF technique as

$$\min_{p_n} \sum_{n=1}^{N} \varpi_n \gamma_n \exp\left(-c_n \left(10 \log_{10}\left(\frac{p_n}{\sigma_n^2}\right) - b_n\right)\right)$$
$$+ V\tilde{G} \tag{39a}$$
$$\text{s.t.} \sum_{n=1}^{N} p_n \|\boldsymbol{w}_{n,\text{ZF}}\|_2^2 \leq P^{\max} \tag{39b}$$
$$p_n \geq \Gamma_{n,\text{req}} \sigma_n^2, \forall n \tag{39c}$$

where

$$\tilde{G} = (a_b - a_s) \left[\frac{1}{\psi} \sum_{n=1}^{N} p_n \|\boldsymbol{w}_{n,\text{ZF}}\|_2^2 + P_{\text{cir}} - E_{\text{hav}}\right]^+$$
$$- a_s \left[\frac{1}{\psi} \sum_{n=1}^{N} p_n \|\boldsymbol{w}_{n,\text{ZF}}\|_2^2 + P_{\text{cir}} - E_{\text{hav}}\right]. \tag{40}$$

Note that the feasibility check of (39) is a linear problem as

$$\text{find } \{p_n\}$$
$$\text{s.t.} \sum_{n=1}^{N} p_n \|\boldsymbol{w}_{n,\text{ZF}}\|_2^2 \leq P^{\max} \tag{41}$$
$$p_n \geq \Gamma_{n,\text{req}} \sigma_n^2, \forall n$$

which is easy to solve.

The corresponding (24a) and (24b) reduce to

$$\gamma_n = \frac{1}{1 + \exp\left(-c_n \left(10 \log_{10}\left(\frac{p_n}{\sigma_n^2}\right) - b_n\right)\right)} \tag{42a}$$

$$\varpi_n = \frac{q_n}{1 + \exp\left(-c_n \left(10 \log_{10}\left(\frac{p_n}{\sigma_n^2}\right) - b_n\right)\right)}. \tag{42b}$$

Based on above analysis, we propose the ZFBF algorithm, and the detailed procedure is summarized in Algorithm 3.

---

**Algorithm 3** ZFBF Algorithm

1: BS sets the iteration index $\tau = 0$, the stop threshold $\epsilon$ and the maximum number of iteration $T^{\max}$, traffic queue backlog $q_n(t), \forall n$.
2: BS obtains a feasible point $\boldsymbol{p}^{(0)}$ via (41).
3: BS updates $\left\{\varpi_n^{(0)}\right\}_{n \in \mathcal{N}}$ and $\left\{\gamma_n^{(0)}\right\}_{n \in \mathcal{N}}$ via (70) and (71), respectively.
4: **repeat**
5: $\quad \tau := \tau + 1$
6: $\quad$ With $\left\{\gamma_n^{(\tau-1)}, \varpi_n^{(\tau-1)}\right\}_{n \in \mathcal{N}}$, BS solves the optimization problem (39) and obtains the power vector $\boldsymbol{p}^{(\tau)}$ for the $\tau$-th iteration.
7: $\quad$ BS updates the following parameters:

$$\gamma_n^{(\tau)} = \frac{1}{1 + \exp\left(-c_n \left(10 \log_{10}\left(\frac{p_n^{(\tau-1)}}{\sigma_n^2}\right) - b_n\right)\right)} \tag{43}$$

$$\varpi_n^{(\tau)} = \frac{q_n}{1 + \exp\left(-c_n \left(10 \log_{10}\left(\frac{p_n^{(\tau-1)}}{\sigma_n^2}\right) - b_n\right)\right)} \tag{44}$$

where $\boldsymbol{\gamma}^{(\tau)} \triangleq \left[\gamma_1^{(\tau)}, \gamma_2^{(\tau)}, \ldots, \gamma_n^{(\tau)}\right]$ and $\boldsymbol{\varpi}^{(\tau)} \triangleq \left[\varpi_1^{(\tau)}, \varpi_2^{(\tau)}, \ldots, \varpi_n^{(\tau)}\right]$.
8: **until** $\frac{\|\boldsymbol{\gamma}^{(\tau)} - \boldsymbol{\gamma}^{(\tau-1)}\|_2}{\|\boldsymbol{\gamma}^{(\tau-1)}\|_2} \leq \epsilon$ and $\frac{\|\boldsymbol{\varpi}^{(\tau)} - \boldsymbol{\varpi}^{(\tau-1)}\|_2}{\|\boldsymbol{\varpi}^{(\tau-1)}\|_2} \leq \epsilon$ or $\tau > T^{\max}$.

---

*Proposition 4:* The proposed ZFBF algorithm converges to a KKT point of (22) with $N_T \geq N$.

*Proof:* The proof is similar to the that of the Proposition 3, and we omit the proof due to the space limitation. □

### D. Complexity Discussion

The ZFBF based optimization problem (39) contains $N$ variables, and the SABF based optimization problem (31) contains $(N_T + 2)N$ variables. The computational complexities of SABF algorithm and ZFBF algorithm are, respectively, $\mathcal{O}\left((N_T + 2)^3 N^3\right)$ and $\mathcal{O}\left(N^3\right)$ via the interior point method [37]. Obviously, the computation complexity for solving (39) is lower than that for solving (27) via the same numerical method.

## V. SIMULATION RESULTS

In this section, we present simulation results to demonstrate the performances of the proposed algorithms. The simulation parameters are specified in Table I.

TABLE I
SIMULATION PARAMETERS SETTING

| Parameters | Values |
|---|---|
| Number of UEs, $N$ | 3 |
| Number of antenna on the BS, $N_T$ | 4 |
| Noise plus background interference power, $\sigma_n^2$ | $10^{-3}$ mW |
| Circuit power cost of the BS, $P_{\text{sp,b}}$ | 115 mW |
| Maximum TX power of each RRH, $P^{\max}$ | 200 mW |
| Minimum SINR requirement, $\Gamma_{n,\text{req}}, u_{\text{avg}}$ | 2 dB, 0.3 |
| Distance between WN and RRH | 10 m |
| Price for buying a unit energy, $a_b$ | 1.2 Cents/mW |
| Price for selling a unit energy, $a_s$ | 1 Cents/mW |
| Forming factors, $b_n, c_n$ | 20, 0.451 |
| Pathloss factor, $\chi$ | 3 |

### A. Iteration of the SABF and ZFBF Algorithms

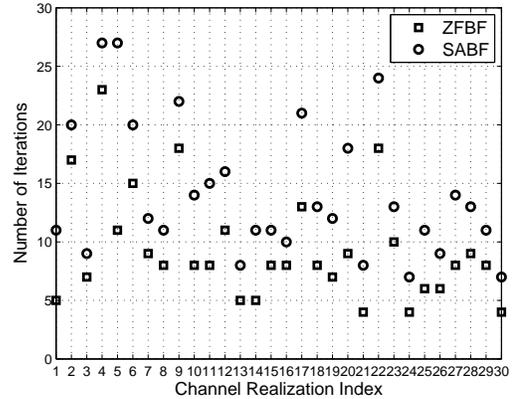

Fig. 2. Iteration numbers for the SABF and ZFBF algorithms, obtained for 30 different channel realizations with control parameter V = 0.001 and initial traffic queue backlog $q_n(0) = 5$.

Figure 2 illustrates the iteration numbers of the SABF algorithm and the ZFBF algorithm obtained for 30 different channel realizations. We select the initial point of the SABF algorithm as the output of the ZFBF algorithm when the ZFBF algorithm is feasible; otherwise, we randomly select a feasible point for the SABF algorithm in the feasible region. As a result, the iteration numbers of the SABF algorithm is counted as the summation of iterations for finding the feasible points via the ZFBF algorithm and the iterations for finding the local optimal solution. We observe that in most of the considered channel realizations, the SABF algorithm and the ZFBF algorithm can converge to the local optimal value within 20 iterations.

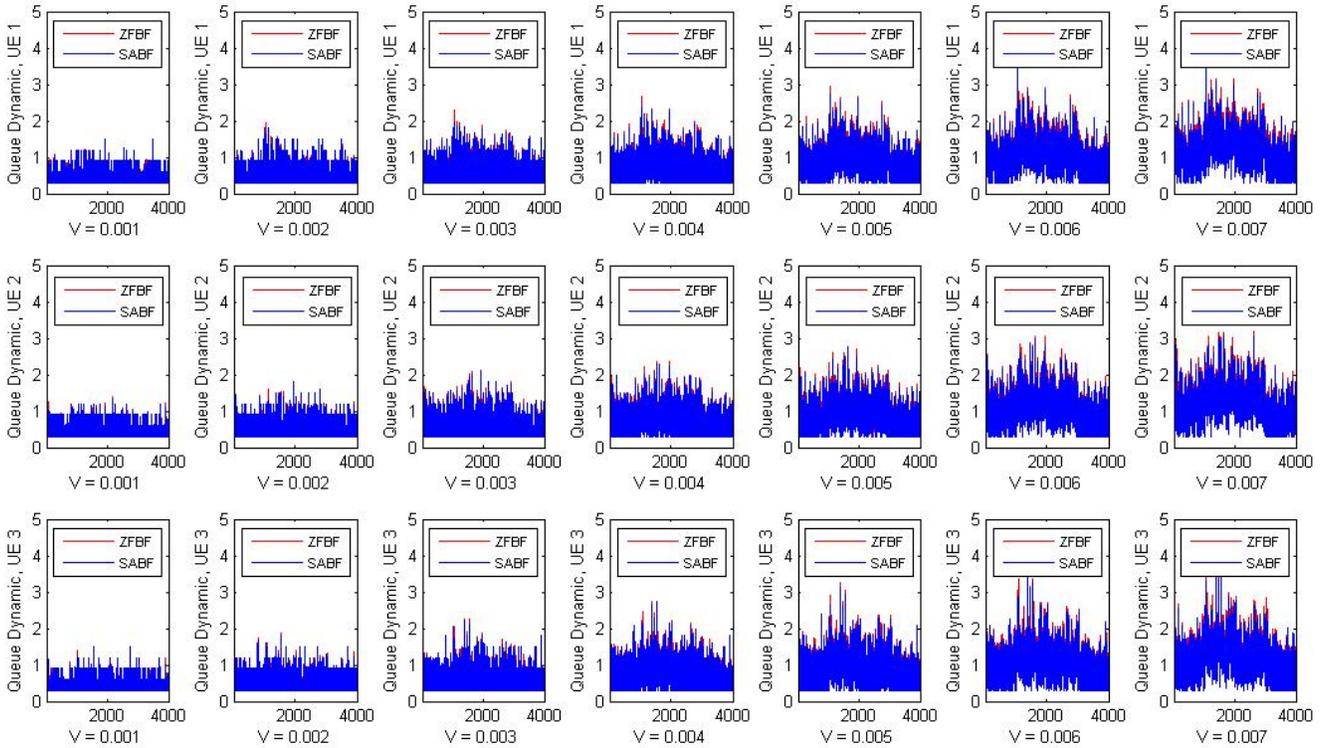

Fig. 3. An illustration of the variations of the backlog of the traffic queues with different control parameters V.

### B. Dynamics of Traffic Queues and Grid Energy Expenditure

Figure 3 illustrates the dynamics of the traffic queue of each WN in 4000 frames. This observation demonstrates that combining the proposed dynamic beamforming framework with the proposed SABF algorithm and ZFBF algorithm can stabilize the traffic queues in the HPC system. We also observe that a larger $V$ leads to a larger backlog of the traffic queue of each WN, which indicates a larger packet delay of each WN according to the Little's law. In other words, the packet delay of each WN increases with the increasing control parameter $V$.

Figure 4 illustrates that the dynamic of the grid energy expenditure over 4000 frames. We consider that the energy arrival varies per 1000 frames. The amount of energy arrival of each frame in the four time durations are 200 mW, 100 mW, 150 mW and 300 mW. The price of selling a unit of harvested energy is 1 Cent/mW, and the price of purchasing a unit of grid energy varies per 500 frames. The prices of purchasing a unit of grid energy for the 8 segments are 1.2 Cents/mW, 1.3 Cents/mW, 1.9 Cents/mW, 1.8 Cents/mW, 1.6 Cents/mW, 1.7 Cents/mW, 1.2 Cents/mW and 1.1 Cents/mW. We observe that the average grid energy consumption fluctuates at the 500-th, 1000-th, 1500-th, 2000-th, 2500-th, 3000-th, 3500-th and 4000-th frame. For the 500-th, 1500-th, 2500-th and 3500-th frame, the fluctuations of the average grid energy consumption is due to the variation of the price of purchasing grid energy. However, the fluctuations of the average grid energy expenditure for the 1000-th, 2000-th, 3000-th and 4000-th frame are two folds: 1) the varying price of purchasing grid energy; 2) the varying amount of harvested energy. We also observe that the BS usually uses the maximum transmission power when the control parameter $V$ is 0.001, and the BS rarely uses the maximum transmission power with $V = 0.007$. This observation indicates that the average transmission power is reduced via increasing $V$; therefore, the proposed Algorithm 1 can effectively control the average transmission power via tuning $V$.

### C. Tradeoff Between Grid Energy Expenditure and Packet Delay

Figures 5 and 6 show the tradeoff between the grid energy expenditure and the delay of WN under different transmission power budgets of the BS and different minimum QoS requirements, respectively. From Figs. 5(a) and 6(a), we observe that the average grid energy expenditure decreases monotonically with the control parameter $V$. The rate of average grid energy expenditure decreasing becomes diminishing as the value of $V$ increases. On the other hand, the packet delay increases with $V$ as shown in Figs. 5(b) and 6(b). Therefore, the operator of the HPC system can tune the control parameter to achieve the

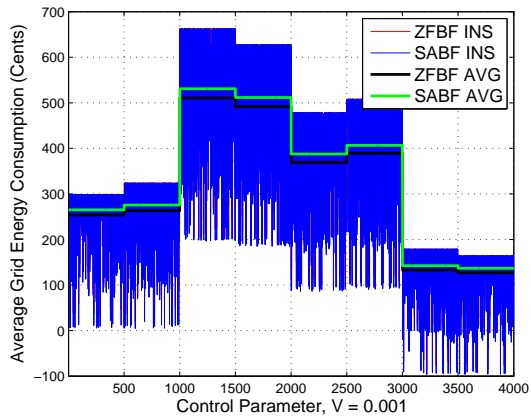

(a) $V = 0.001$

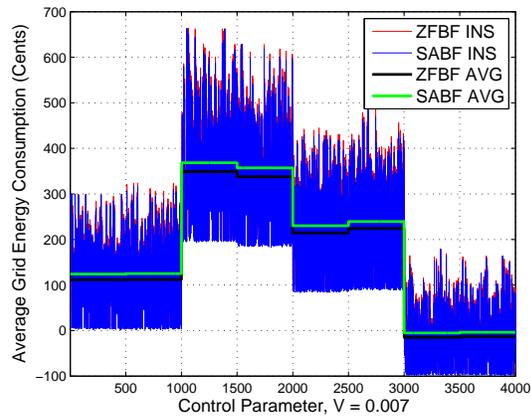

(b) $V = 0.007$

Fig. 4. An illustration of the variations of the grid energy expenditure.

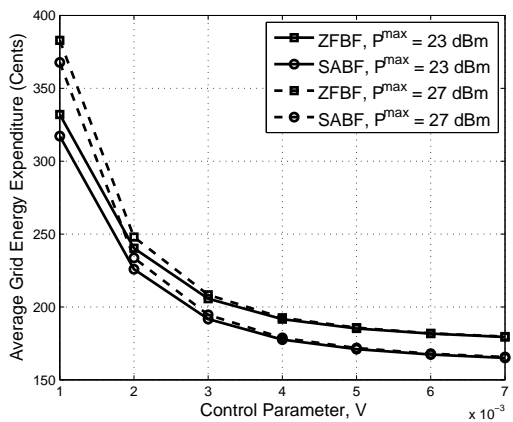

(a)

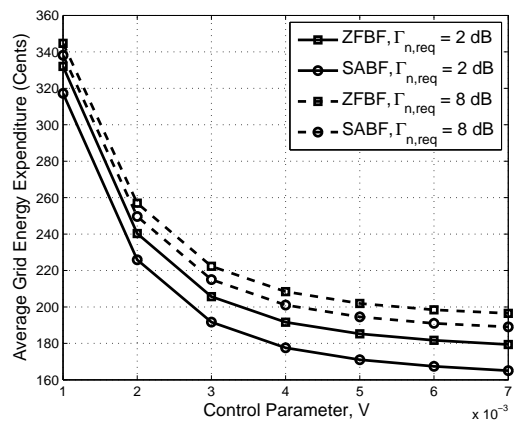

(a)

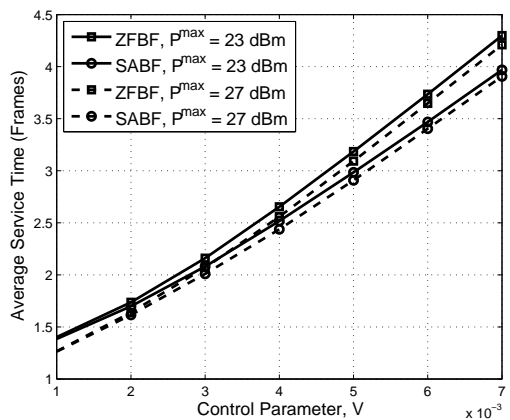

(b)

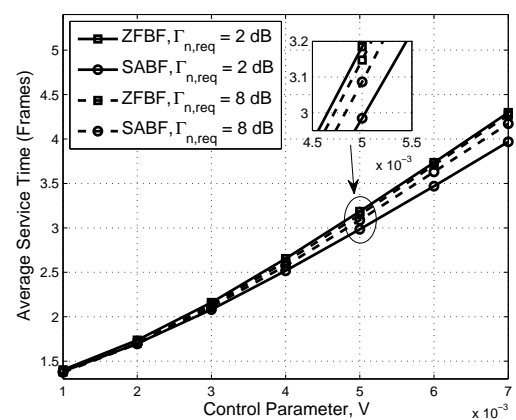

(b)

Fig. 5. Tradeoff between the grid energy expenditure and the packet delay of WN with different power budgets.

Fig. 6. Tradeoff between the grid energy expenditure and the packet delay of WN with different SINR requirements.

target grid energy expenditure at the expense of the packet delay of the UEs. Besides, the SABF algorithm performs better than the ZFBF algorithm in the average grid energy expenditure and the packet delay under different parameter settings. As shown in Fig. 5(a), the performance gain in average grid energy expenditure of the case $P^{\max} = 27$ dBm diminishes compared with the case $P^{\max} = 23$ dBm when the value of $V$ is greater than 0.005. However, the packet delay of the case $P^{\max} = 27$ dBm can still have around 1% improvement than the case $P^{\max} = 23$ dBm. This is due to the fact that the BS with more transmission power budget can enhance the sigmoidal packet departure rate when the channel condition is severe. As a result, the packet delay with $P^{\max} = 27$ dBm can reduce the packet delay compared with the case $P^{\max} = 23$ dBm.

*D. The Impact of SINR Requirements on Grid Energy Expenditure and Packet Delay*

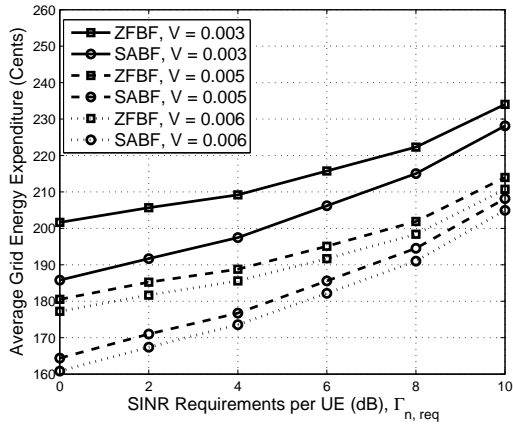

(a) Grid energy expenditure v.s. control parameter $V$.

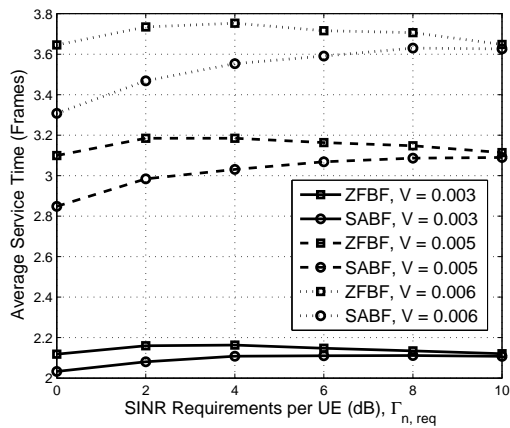

(b) Delay v.s. control parameter $V$.

Fig. 7. The impact of SINR requirements on the grid energy expenditure and the delay of WN.

Figure 7 illustrates the impact of SINR on the grid energy expenditure and the delay of WN. As shown in Fig. 7(a), the grid energy expenditure increases monotonically with the instantaneous SINR requirements for the SABF algorithm and the ZFBF algorithm. This can be explained by the fact that the more power needs to be consumed in order to satisfy the instantaneous SINR requirements. From Fig. 7(b), we also observe that the packet delay of the proposed SABF algorithm performs better than that of the ZFBF algorithm. Increasing the instantaneous SINR requirements leads to the gap of the packet delay diminishes between the SABF algorithm and the ZFBF algorithm. This is due to the fact that the ZFBF algorithm can obtain a near optimal solution in the high SINR region, for example, $\Gamma_{n,\text{req}} \geq 10$ dB in our setting.

## VI. CONCLUSION

We developed the cross-layer beamforming in the HPC systems with HUT strategy, where the BS can purchase electricity from the smart grid if the harvested energy is insufficient and sell the surplus harvested energy to generate revenue. We leverage a stochastic optimization model to formulate the long-term grid energy expenditure minimization problem. Reformulating the stochastic optimization model as a per-frame GEWPR minimization problem, we revealed tradeoff between the long-term grid energy expenditure and the packet delay. For example, setting a large value of control parameter $V$, the grid expenditure can be reduced at the expense of large packet delay of each WN. Therefore, the operators can set the control parameter according to the arrival rate of the renewable energy. Due to the NP-hardness of the per-frame GEWPR problem, two suboptimal algorithms, namely SABF algorithm and ZFBF algorithm, were proposed. The SABF algorithm outperforms the ZFBF algorithm in both the long-term grid energy expenditure and the packet delay at the expense of higher computational complexity.

## APPENDIX A
## PROOF OF LEMMA 1

Substituting (10) into (16), we have

$$\Delta\left(\boldsymbol{q}\left(t\right)\right)$$
$$= \frac{1}{2}\sum_{n=1}^{N}\mathbb{E}\left[\left(q_n^2(t+1) - q_n^2(t)\right)|\boldsymbol{q}(t)\right]$$
$$\overset{(a)}{\leq} \sum_{n=1}^{N}\mathbb{E}\left[\frac{1}{2}\left(q_n^2(t) + U_n^2(t) + U_{n,\text{req}}^2(t)\right)\right]$$
$$+ \sum_{n=1}^{N}\mathbb{E}\left[q_n(t)\left(U_{n,\text{req}}(t) - U_n(t)\right) - \frac{1}{2}q_n^2(t)|\boldsymbol{q}(t)\right]$$
$$\overset{(b)}{\leq} N + \sum_{n=1}^{N}\mathbb{E}\left[q_n(t)\left(U_{n,\text{req}}(t) - U_n(t)\right)|\boldsymbol{q}(t)\right] \quad (45)$$

where (a) is due to the fact $\left([a-b,0]^+ + c\right)^2 \leq a^2 + b^2 + c^2 + 2a(c-b)$ with $a,b,c \geq 0$, and (b) is due to the fact that the upper bound of $\frac{1}{2}\left(U_n^2(t) + U_{n,\text{req}}^2(t)\right)$ is one from the the definition in (9). Submitting the penalty term $V\mathbb{E}\left[G(t)|\boldsymbol{q}(t)\right]$ to both sides of (45), we obtain (17).

## APPENDIX B
## PROOF OF PROPOSITION 1

The major steps follow the proof of [28, Theorem 4.2], and we will only sketch the proof for our formulated optimization problem. The optimization problem (18) is NP-hard; therefore, we are motivated to seek the suboptimal solutions within the feasible region. Given $\{\boldsymbol{h}_n(t), q_n(t)\}_{n \in \mathcal{N}}$, we defined the RHS of (17) as

$$f\left(\{\boldsymbol{w}_n(t)\}_{n \in \mathcal{N}}\right) = N + V \mathbb{E}\left[G(t) | \boldsymbol{q}(t)\right] \\ + \sum_{n=1}^{N} q_n(t) \mathbb{E}\left[U_{n,\text{req}}(t) - U_n(t) | \boldsymbol{q}(t)\right] \quad (46)$$

where

$$G(t) = \tilde{G}\left(\{\boldsymbol{w}_n(t)\}_{n \in \mathcal{N}}\right) \quad (47)$$
$$U_n(t) = \tilde{U}_n\left(\{\boldsymbol{w}_n(t)\}_{n \in \mathcal{N}}\right). \quad (48)$$

Generally, the suboptimal solutions lead to a gap between the optimal value of (46) and the suboptimal value of (46), and this gap is generally difficult to obtain. Given $\{\boldsymbol{h}_n(t), q_n(t)\}_{n \in \mathcal{N}}$, we denote the suboptimal solutions and the optimal solution, respectively, as $\{\boldsymbol{w}_n^*(t)\}_{n \in \mathcal{N}}$ and $\{\boldsymbol{w}_n^{\text{OPT}}(t)\}_{n \in \mathcal{N}}$. Moreover, we have $f\left(\{\boldsymbol{w}_n^{\text{OPT}}(t)\}_{n \in \mathcal{N}}\right) \leq f\left(\{\boldsymbol{w}_n^*(t)\}_{n \in \mathcal{N}}\right)$. Therefore, the term $\aleph$ equals to $f\left(\{\boldsymbol{w}_n^{\text{OPT}}(t)\}_{n \in \mathcal{N}}\right) - f\left(\{\boldsymbol{w}_n^*(t)\}_{n \in \mathcal{N}}\right)$, which is smaller than or equals to zero. Thus, the upper bound of RHS of (17) is denoted by

$$\Delta(\boldsymbol{q}(t)) + V \mathbb{E}\left[G(t) | \boldsymbol{q}(t)\right] \\ \leq \aleph + f\left(\{\boldsymbol{w}_n^*(t)\}_{n \in \mathcal{N}}\right) \\ \leq f\left(\{\boldsymbol{w}_n^*(t)\}_{n \in \mathcal{N}}\right). \quad (49)$$

When $[u_{1,\text{req}}, \ldots, u_{N,\text{req}}] \in \boldsymbol{\lambda} + \epsilon$ and the channel coefficients $\{\boldsymbol{h}_n\}_{n \in \mathcal{N}}$ are independent identically distributed over different frames, we have the following two inequalities

$$\mathbb{E}\left[G^*(t) | \boldsymbol{q}(t)\right] \leq G^{\text{SUBOPT}} + \delta \quad (50)$$
$$\mathbb{E}\left[U_n^*(t) | \boldsymbol{q}(t)\right] \geq u_{n,\text{req}} + \epsilon \quad (51)$$

where $G^{\text{SUBOPT}}$ is the maximum suboptimal value of (14). Here, the value of $\delta$ is equal to zero or arbitrarily close to zero [28, Appendix 4.A].

Although it is possible to stabilize the traffic queues when $[u_{1,\text{req}}, \ldots, u_{N,\text{req}}]$ is outside the capacity region achieved by the suboptimal solutions $\boldsymbol{\lambda}$, the computation complexity is extremely high and impractical for the real systems. Once $[u_{1,\text{req}}, \ldots, u_{N,\text{req}}]$ is outside the capacity region $\boldsymbol{\lambda}$, we can improve the maximum transmission power in order to stabilize the traffic queues at the design stage of the HPC system.

Besides, the expected grid energy expenditure is lower bounded as

$$\mathbb{E}\left[G(t)\right] \geq G^{\min} \quad (52)$$

where $G^{\min}$ is the smallest value of grid energy expenditure. This constraint is reasonable because any physical parameter is bounded below and above in practical systems.

Substituting (50) and (51) into (49) and letting $\delta \to 0$, we obtain

$$\Delta(\boldsymbol{q}(t)) + V \mathbb{E}\left[G(t) | \boldsymbol{q}(t)\right] \\ \leq N + V G^{\text{SUBOPT}} - \epsilon \sum_{n=1}^{N} q_n(t). \quad (53)$$

1) In order to prove the first part, we make some algebraic manipulation of (53) as

$$\Delta(\boldsymbol{q}(t)) \\ \leq N + V \left(G^{\text{SUBOPT}} - \mathbb{E}\left[G(t) | \boldsymbol{q}(t)\right]\right) - \epsilon \sum_{n=1}^{N} q_n(t) \\ \leq N + V \left(G^{\text{SUBOPT}} - \mathbb{E}\left[G(t) | \boldsymbol{q}(t)\right]\right) \quad (54)$$

Taking the iterated expectation over $\boldsymbol{q}(t)$ of (54), we have

$$\frac{1}{2}\mathbb{E}\left[\|\boldsymbol{q}(t+1)\|_2^2\right] - \frac{1}{2}\mathbb{E}\left[\|\boldsymbol{q}(t)\|_2^2\right] \\ \leq N + V \left(G^{\text{SUBOPT}} - \mathbb{E}\left[G(t)\right]\right) \\ \leq N + V \left(G^{\text{SUBOPT}} - G^{\min}\right). \quad (55)$$

Using the telescoping sum over $\{0, 1, 2, \ldots, t-1\}$, we obtain

$$\mathbb{E}\left[\|\boldsymbol{q}(t)\|_2^2\right] \\ \leq 2t\left(N + V\left(G^{\text{SUBOPT}} - G^{\min}\right)\right) + \mathbb{E}\left[\|\boldsymbol{q}(0)\|_2^2\right]. \quad (56)$$

On the other hand, we have

$$\text{Var}\left[|q_n(t)|\right] = \mathbb{E}\left[|q_n(t)|^2\right] - \mathbb{E}^2\left[|q_n(t)|\right] \geq 0 \quad (57)$$
$$\mathbb{E}\left[|q_n(t)|^2\right] \leq \mathbb{E}\left[\|\boldsymbol{q}(t)\|_2^2\right]. \quad (58)$$

Substituting (57) and (58) into (56) and performing some algebraic operations, we can achieve

$$\mathbb{E}\left[|q_n(t)|\right] \\ \leq \sqrt{2t\left(N + V\left(G^{\text{SUBOPT}} - G^{\min}\right)\right) + \mathbb{E}\left[\|\boldsymbol{q}(0)\|_2^2\right]}. \quad (59)$$

Dividing (59) by $t$ and letting $t \to \infty$, we obtain

$$\lim_{t \to \infty} \frac{1}{t} \mathbb{E}\left[|q_n(t)|\right] = 0. \quad (60)$$

Hence, we conclude that the virtual queue is mean rate stable.

2) Taking iterated expectation over $\boldsymbol{q}(t)$ of (53) and using the telescoping sum over $\{0, 1, 2, \ldots, t-1\}$, we have

$$\frac{1}{2}\mathbb{E}\left[\|\boldsymbol{q}(t)\|_2^2\right] - \frac{1}{2}\mathbb{E}\left[\|\boldsymbol{q}(0)\|_2^2\right] + V \sum_{\tau=0}^{t-1} \mathbb{E}\left[G(\tau)\right] \\ \leq t\left(N + V G^{\text{SUBOPT}}\right) - \epsilon \sum_{\tau=0}^{t-1} \sum_{n=1}^{N} \mathbb{E}\left[q_n(\tau)\right]. \quad (61)$$

Dividing (61) by $Vt$ and making some algebraic manipulations, we obtain

$$\frac{1}{t} \sum_{\tau=0}^{t-1} \mathbb{E}\left[G(\tau)\right] \\ \leq \frac{N}{V} + G^{\text{SUBOPT}} - \frac{\epsilon}{Vt} \sum_{\tau=0}^{t-1} \sum_{n=1}^{N} \mathbb{E}\left[q_n(\tau)\right] + \frac{1}{2Vt} \mathbb{E}\left[\|\boldsymbol{q}(0)\|_2^2\right] \\ \leq \frac{N}{V} + G^{\text{SUBOPT}} + \frac{1}{2Vt} \mathbb{E}\left[\|\boldsymbol{q}(0)\|_2^2\right]. \quad (62)$$

Letting $t \to \infty$, we prove (19) due to the fact that $\mathbb{E}\left[\|\boldsymbol{q}(t)\|_2^2\right] < \infty$.

3) From (61), we have

$$\epsilon \sum_{\tau=0}^{t-1} \sum_{n=1}^{N} \mathbb{E}\left[q_n(\tau)\right] - \frac{1}{2}\mathbb{E}\left[\|\boldsymbol{q}(0)\|_2^2\right] + V \sum_{\tau=0}^{t-1} \mathbb{E}\left[G(\tau)\right]$$
$$\leq t\left(N + VG^{\text{SUBOPT}}\right). \tag{63}$$

Substituting (52) into (63) and performing some algebraic manipulations, we have

$$\frac{1}{t} \sum_{\tau=0}^{t-1} \sum_{n=1}^{N} \mathbb{E}\left[q_n(\tau)\right]$$
$$\leq \frac{N + V\left(G^{\text{SUBOPT}} - G^{\min}\right)}{\epsilon} + \frac{1}{2t}\mathbb{E}\left[\|\boldsymbol{q}(0)\|_2^2\right]. \tag{64}$$

Letting $t \to \infty$, we obtain (20) due to the fact that $\mathbb{E}\left[\|\boldsymbol{q}(t)\|_2^2\right] < \infty$.

## APPENDIX C
## PROOF OF PROPOSITION 2

Introducing a set of auxiliary variables $\gamma_n > 0$, $n \in \mathcal{N}$, we obtain an equivalent form of the optimization problem (22) as

$$\min_{\boldsymbol{w}_n, \alpha_n, \gamma_n} VG - \sum_{n=1}^{N} q_n \gamma_n \tag{65a}$$

$$\text{s.t.} \sum_{n=1}^{N} \|\boldsymbol{w}_n\|_2^2 \leq P^{\max} \tag{65b}$$

$$\text{SINR}_n \geq \Gamma_{n,\text{req}}, \forall n \tag{65c}$$

$$\text{SINR}_n \geq \alpha_n, \forall n \tag{65d}$$

$$\gamma_n \left(1 + \exp\left(-c_n \left(10 \log_{10}(\alpha_n) - b_n\right)\right)\right) \leq 1, \forall n. \tag{65e}$$

The equivalence of (65) and (22) can be easily justified. We note that all constraints in (65e) are active at the optimum. Otherwise, the value of the objective function (65a) is decreased by increasing $\gamma_n$.

The Lagrangian of (65) is denoted as

$$L = VG - \sum_{n=1}^{N} q_n \gamma_n + \varsigma \left(\sum_{n=1}^{N} \|\boldsymbol{w}_n\|_2^2 - P^{\max}\right)$$
$$+ \sum_{n=1}^{N} \varpi_n \left(\gamma_n \left(1 + \exp\left(-c_n \left(10 \log_{10}(\alpha_n) - b_n\right)\right)\right) - 1\right)$$
$$+ \sum_{n=1}^{N} \varrho_n \left(\Gamma_{n,\text{req}} - \text{SINR}_n\right)$$
$$+ \sum_{n=1}^{N} \zeta_n \left(\alpha_n - \text{SINR}_n\right) \tag{66}$$

where $\varsigma$ and $(\varpi_n, \varrho_n, \zeta_n)_{n \in \mathcal{N}}$ are non-negative Lagrange multipliers.

The KKT conditions related to the proof are listed as

$$\varpi_n \frac{\partial L}{\partial \varpi_n} \tag{67}$$
$$= \varpi_n \left(\gamma_n \left(1 + \exp\left(-c_n \left(10 \log_{10}(\alpha_n) - b_n\right)\right)\right) - 1\right) = 0, \forall n$$

$$\frac{\partial L}{\partial \gamma_n} \tag{68}$$
$$= \varpi_n \left(1 + \exp\left(-c_n \left(10 \log_{10}(\alpha_n) - b_n\right)\right)\right) - q_n = 0, \forall n$$

$$\varpi_n \geq 0, \forall n. \tag{69}$$

From (68), we obtain the expression for $\varpi_n$ as

$$\varpi_n = \frac{q_n}{1 + \exp\left(-c_n \left(10 \log_{10}(\alpha_n) - b_n\right)\right)} > 0. \tag{70}$$

Hence, we can obtain the expression of $\gamma_n$ from (67) as

$$\gamma_n = \frac{1}{1 + \exp\left(-c_n \left(10 \log_{10}(\alpha_n) - b_n\right)\right)}. \tag{71}$$

With (70) and (71), we obtain the following optimization problem

$$\min_{\boldsymbol{w}_n, \alpha_n} \sum_{n=1}^{N} \varpi_n \left(\gamma_n \left(1 + \exp\left(-c_n \left(10 \log_{10}(\alpha_n) - b_n\right)\right)\right) - 1\right)$$
$$+ VG - \sum_{n=1}^{N} q_n \gamma_n \tag{72a}$$

$$\text{s.t.} \sum_{n=1}^{N} \|\boldsymbol{w}_n\|_2^2 \leq P^{\max} \tag{72b}$$

$$\text{SINR}_n \geq \Gamma_{n,\text{req}}, \forall n \tag{72c}$$

$$\text{SINR}_n \geq \alpha_n, \forall n. \tag{72d}$$

With (70) and (71), the optimization problem (72) has the same KKT conditions with the optimization problem (22). Dropping the constant in the objective function (72a), we obtain the optimization problem (25).

## APPENDIX D
## ACTIVENESS OF CONSTRAINTS IN (27e)

With (23), we can obtain the $\beta_n \sqrt{\alpha_n^*} \geq \beta_n \sqrt{\Gamma_{n,\text{req}}}$, where $\alpha_n^*$ represents a local optimal value of $\alpha_n$. Thus, we only discuss the relation between (27d) and (27e) in this proof.

Suppose some of the constraints in (27e) are inactive. We denote the set of indices of inactive constraints as

$$\mathcal{I} = \left\{ i \,\Bigg|\, \sqrt{\sigma_i^2 + \sum_{m=1, m \neq i}^{N} \left|\boldsymbol{h}_i^{\text{H}} \boldsymbol{w}_m\right|^2} < \beta_i \right\}. \tag{73}$$

Then, there must be a positive constant $\mu > 1$ such that $\sqrt{\sigma_i^2 + \sum_{m=1, m \neq i}^{N} \left|\boldsymbol{h}_i^{\text{H}} \boldsymbol{w}_m\right|^2} = \hat{\beta}_i$ and $\hat{\beta}_i = \frac{\beta_i}{\mu}$, $i \in \mathcal{I}$. To keep the corresponding constraint in (27d) unchanged, we set $\hat{\alpha}_i = \mu^2 \alpha_i^*$ such that $\beta_i \sqrt{\alpha_i^*} = \hat{\beta}_i \sqrt{\hat{\alpha}_i}$. Obviously, $\hat{\alpha}_i = \mu^2 \alpha_i^* > \alpha_i^*$ for $\mu > 1$. Then, the term $\hat{\alpha}_i$ leads to a smaller value of (27a), which contradicts the fact that the term $\alpha_i^*$ obtains its local optimal value. As a result, we conclude that constraints in (27e) are always active.

# APPENDIX E
# PROOF OF PROPOSITION 3

The proposed Algorithm 2 uses successive approximation procedures. Let $\mathcal{F}^{(\tau)}$ and $\mathcal{OBJ}^{(\tau)}$ respectively denote the approximated feasible region and the optimal value of the objective function of the optimization problem (27) in the $\tau$-th iteration. From Line 2-3 of Algorithm 2, we obtain that the point $(\boldsymbol{w}_n^{(\tau-1)}, \alpha_n^{(\tau-1)}, \beta_n^{(\tau-1)}, \varpi_n^{(\tau-1)}, \gamma_n^{(\tau-1)})$ is in the feasible region $\mathcal{F}^{(\tau)}$. Therefore, we conclude that

$$\mathcal{OBJ}^{(\tau)} \leq \mathcal{OBJ}^{(\tau-1)} \tag{74}$$

because the point $(\boldsymbol{w}_n^{(\tau)}, \alpha_n^{(\tau)}, \beta_n^{(\tau)}, \varpi_n^{(\tau)}, \gamma_n^{(\tau)})$ is the optimal point in the $\tau$-th iteration according to Line 6 of Algorithm 2. On the other hand, the transmission power of the BS is limited by $P^{\max}$, which indicates that the value of the objective function in (27) is lower bounded. Hence, the generated sequence $(\boldsymbol{w}_n^{(\tau)}, \alpha_n^{(\tau)}, \beta_n^{(\tau)}, \varpi_n^{(\tau)}, \gamma_n^{(\tau)})$ converges as $\tau \to \infty$.

Because the objection function (27a) is strictly convex, we can prove the convergent point $(\boldsymbol{w}_n^{(\infty)}, \alpha_n^{(\infty)}, \beta_n^{(\infty)}, \varpi_n^{(\infty)}, \gamma_n^{(\infty)})$ is a KKT point of the optimization problem (22) via similar arguments in [38, Proposition 3.2].